\documentclass{elsart}
\usepackage{graphicx,amssymb}
\usepackage{natbib}

\def\astrobj#1{#1}
\journal{New Astronomy}

\begin{document}

\begin{frontmatter}

\title{COOLING FLOW BULK MOTION CORRECTIONS TO THE SUNYAEV-ZEL'DOVICH EFFECT}

\author{P.\,M.\,Koch\thanksref{cor}},
\thanks[cor]{Corresponding author: pmkoch@physik.unizh.ch}\author{Ph.\,Jetzer}, \author{D.\,Puy}

\address{Institute of Theoretical Physics, University of Z\"urich,
         Winterthurerstrasse 190, CH-8057 Z\"urich, Switzerland}

\begin{abstract}
We study the influence of converging cooling flow bulk motions
on the Sunyaev-Zel'dovich (SZ) effect.
To that purpose we derive a modified Kompaneets equation
which takes into account the contribution
of the accelerated electron media of the cooling flow
inside the cluster frame. The additional term
is different from the usual kinematic SZ-effect, which
depends linearly on the velocity, whereas the contribution 
described here is quadratic in the macroscopic electron fluid velocity, 
as measured in the cluster frame. 
For clusters with a large cooling flow mass deposition rate and/or a small
central electron density, it turns out that this effect
becomes relevant.
\end{abstract}
 
\end{frontmatter}

\section{Introduction}
\label{intro}

The SZ-effect is becoming more and more an important
astrophysical tool thanks to the rapid progress of the observational
techniques, which allow increasingly precise measurements. It has 
thus become relevant to study further corrections to it, such as relativistic 
effects \citep{re95}, the shape of the
galaxy cluster and its finite extension or a polytropic temperature profile
(see e.g. \citet{pu00}), the presence of cooling flows \citep{sc91,ma00}, 
corrections induced by halo rotation \citep{co01} or even by
Brillouin scattering \citep{sm02} and the influence
of early galactic winds \citep{ma01}.
These additional effects are of different relevance and often depend on the
specific values of the parameters which describe 
a given cluster. Generally, they range from the percent level up to
even 20-30\% and accordingly in the subsequent determination of the Hubble 
constant. 
Taking into account such corrections is on one side
relevant when determining the Hubble constant via the SZ-effect, and on the
other hand they could be interesting as a tool to study the detailed 
structure of the cluster itself.\\

Following these lines we study here another effect, which 
has not yet been taken into account: The possible influence on the SZ-effect
of the cooling flow 
bulk motion of the electron media inside the cluster frame. 
Indeed, one expects such 
a motion to be present in cooling flows, for 
which in some clusters there is evidence in the central regions
\citep{al93,fa91,fa94,fa97}.
Recently, \citet{db02} investigated the prospects 
to detect gas bulk velocities in clusters of galaxies through
the kinematic SZ-effect, which depends linearly on the 
velocity component along the line of sight.
Bulk motions along a given line of sight will contribute to the
kinematic SZ term as long as their averaged velocity, in the observer frame,
does not vanish.
In the case of cooling flow bulk motion the averaged velocity, in the
cluster frame, along a
given line of sight vanishes in good approximation, since we assume spherical
symmetric infall. Thus, the cooling flow bulk motion will not contribute
as such to the kinematic SZ-effect.
Indeed, the effect we  consider depends quadratically on 
the velocity and clearly, the averaged quadratic velocity does not
vanish along a line of sight in the cluster frame.  
As we will see, the considered effect usually turns out to be small, since
the cooling flow
bulk motion velocities are rather small, unless in the very central regions
of a cooling flow clusters, where the cooling 
flow approaches a sonic radius and changes
from the subsonic to the supersonic regime. Nonetheless, in some favourable
cases it might be of the order of some percent of the thermal SZ-effect.\\

The aim of
this paper is to examine the influence on the thermal SZ-effect 
of the moving electron media inside a cooling
flow region of a galaxy cluster. 
As it will be seen in more detail later, 
this effect has clearly to be distinguished from the kinematic SZ-effect,
which takes into account the motion of the whole cluster or a fraction
of it along a given line of sight.
In fact, the standard thermal SZ-effect describes the frequency 
dependent intensity change of the 
Cosmic Microwave Background (CMB)
photons by inverse Compton scattering off the hot 
intracluster plasma (electrons), where the electrons are
supposed to be static scatterers.\\

The paper is organised as follows: In section 2 we 
briefly outline the dynamics of a homogeneous steady-state cooling flow
model in order to get the velocity profile of the bulk motion
as well as the corresponding electron density distribution.
In section 3 we derive the modification to the standard 
Kompaneets equation 
due to the inclusion of bulk motion of the 
scatterers inside the cluster frame
and in section 4 we present and discuss our results. 
In section 5 we give a summary and an outlook.

\section{Velocity profile for a homogeneous cooling flow model}

\begin{figure}[b]
\begin{center}
\includegraphics[scale=0.5]{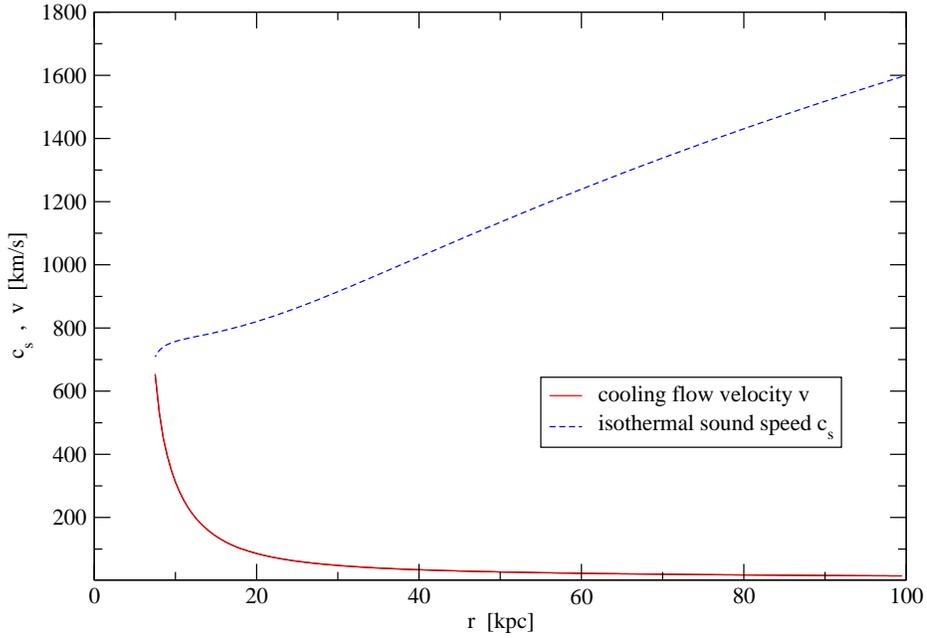}
\caption{\label{cfdynamics}Cooling flow velocity $|v(r)|$ and isothermal sound speed $c_s(r)$ as a function of radius. The mass 
deposition rate is $\dot{m}=-300\,M_{\odot}\,yr^{-1}$, $r_{cool}= 100\,kpc$.
The initial values for the integration are $v(r_{cool})=15\,km/s$ and $c_s(r_{cool})=1600\,km/s$.}
\end{center}
\end{figure}

\begin{figure}
\begin{center}
\includegraphics[scale=0.5]{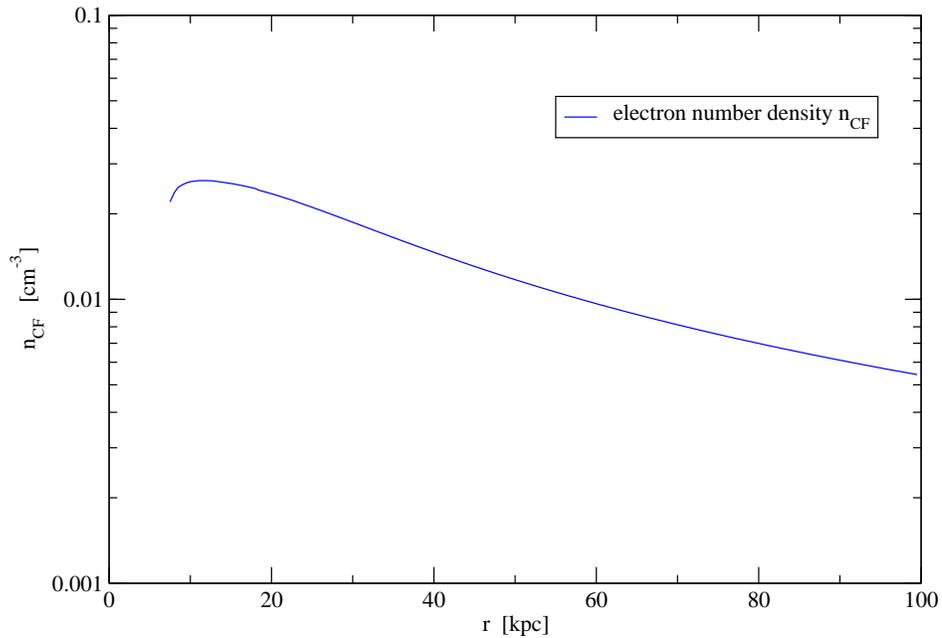}
\caption{\label{cfdensity}Cooling flow electron number density $n_{CF}(r)$ as a function of radius. Cooling flow parameters as
adopted in Fig.\ref{cfdynamics}.}
\end{center}
\end{figure}

As an example to get a velocity profile which describes 
the bulk motion,   
we consider a homogeneous steady-state cooling flow model 
where the mass deposition rate $\dot{m}$ is 
constant, negative and enters as a parameter in the model. 
No mass drops out of the flow. 
We neglect the possible influence of magnetic fields, rotation and 
viscosity. In this context,
the cluster is expected to be in a relaxed state, so 
that hydrostatic equilibrium allows us to use an isothermal $\beta$-model \citep{sa88}.
For spherical symmetry, the cooling flow can thus be 
described by a set of Euler equations. 
Mass, momentum and energy conservation read \citep{mb78,fa84,ws87a,ws87b,sa88}:
\begin{equation}
\left.
\begin{array}{l}
 4\pi r^2\rho_{\scriptscriptstyle CF}v  =  \dot{m}  \\ 
 v\frac{dv}{dr}+\frac{1}{\rho_{\scriptscriptstyle CF}}\frac{dP}{dr}+ \frac{G\mathcal{M}(r)}{r^2}  =  0  
 \label{cf1}    \\
 v\frac{dE}{dr}-\frac{P\,v}{\rho_{\scriptscriptstyle CF}^2}\frac{d\rho_{\scriptscriptstyle CF}}{dr}  =  -\Lambda
 \rho_{\scriptscriptstyle CF},
\end{array}
\right\}
\end{equation}
where $\rho_{\scriptscriptstyle CF}(r)$ and $P(r)$ are the gas density and pressure, respectively, in the 
cooling 
flow and $v(r)<0$ is the velocity of the inward directed cooling flow. $\mathcal{M}(r)$ is the gravitating 
cluster mass inside the radius $r$, discussed earlier in the literature by the above authors, with the dark matter ($DM$) 
density profile as follows:
\begin{equation}
\rho_{DM}(r)=\frac{\rho_0}{1+(r/r_c)^2},
\end{equation}
where the central cluster density $\rho_0=1.8\cdot 10^{-25}g\,cm^{-3}$ and the core radius $r_c=250\,kpc$ describe
the profile of the cluster mass.
As usual, we assume that the cooling flow makes no significant contribution
to the cluster mass density and, that the gas self-gravity can be neglected. The internal energy is 
$E(r)=\frac{3}{2}\theta(r)$, with the temperature parameter $\theta$ which defines the square of the 
isothermal sound
speed $c_s$:
\begin{equation}
\theta(r):=c_s^2(r)=\frac{kT_{\scriptscriptstyle e,CF}(r)}{\mu m_p}, \label{cf2}
\end{equation}
where $\mu\approx 0.63$ is the mean molecular weight, $m_p$ the proton mass and 
$T_{\scriptscriptstyle e,CF}$ 
the electron temperature in the cooling
flow. $\Lambda(\theta)$ defines the radiative optically thin cooling function for a low density plasma,
which is defined so that $\Lambda\rho_{\scriptscriptstyle CF}^2$ is the cooling rate per unit volume in the gas.
We take an analytical fit which has 
been used in earlier works \citep{sw87,ma00}:  
\begin{eqnarray}
& & \frac{\Lambda(\theta)}{10^{-22}erg\,cm^3\,s^{-1}}=4.7\cdot\exp\left[-\left(\frac{T}{3.5\cdot10^5K}\right)
^{4.5}\right]\nonumber\\
&+& 0.313\cdot T^{0.08}\cdot\exp\left[-\left(\frac{T}{3.0\cdot10^6K}\right)^{4.4}\right]\nonumber\\
&+& 6.42\cdot T^{-0.2}\cdot\exp\left[-\left(\frac{T}{2.1\cdot10^7K}\right)^{4.0}\right]\nonumber\\
&+& 0.000439\cdot T^{0.35}.                                                        \label{cf2.1}
\end{eqnarray}
Eliminating the density $\rho_{\scriptscriptstyle CF}$ with the mass conservation equation,  
the Euler system in (\ref{cf1}) leads then to a system of two ordinary coupled differential 
equations for the cooling flow velocity $v(r)$ and the square of the isothermal sound speed $\theta(r)$ \citep{mb78}:
\begin{equation}
\left.\begin{array}{l}
  \frac{dv}{dr}=v\left(3G\mathcal{M}-10r\theta+\frac{\dot{m}}{2\pi}\frac{\Lambda(\theta)}{v^2}\right)\bigg 
   /\bigg(r^2(5\theta-3v^2)\bigg)                       \\
  \frac{d\theta}{dr}=2\left(\theta(2v^2r-G\mathcal{M})-(v^2-\theta)\frac{\dot{m}}{4\pi}\frac{\Lambda(\theta)}
{v^2}\right)\bigg  / \bigg(r^2(5\theta-3v^2)\bigg).                         \label{cf3}
\end{array}\right\} 
\end{equation}
To integrate this system we have to take into account that both equations have singularities at the sonic radius 
$r_s$, where $5\theta(r_s)=3v^2(r_s)$ and where the cooling flow undergoes a transsonic transition. The cooling 
flow region is limited by the cooling radius $r_{cool}\approx 100-150\,kpc$.\\
We stress that our goal is not to develop a sophisticated cooling flow model, but to get an idea of how the cooling
flow bulk motion contributes to the SZ-effect. We, therefore, do not attempt to find solutions to the system of
Eqs.(\ref{cf3}) for 
the innermost supersonic region, $r<r_s$, where we expect shocks to be important. We then avoid the search of critical
values\footnote[1]{A transition from subsonic to supersonic occurs, if the numerators and denominators in the system of 
Eqs.(\ref{cf3})
vanish at $r_s$. Since the expressions are indeterminate, they have to be replaced by nonsingular expressions by the
Bernoulli-de l'H\^opital's rule, which will give initial 
values for Eq.(\ref{cf3}) at $r_s$ if one assumes a temperature at the
sonic radius.}
at $r_s$ which would have to be matched to hydrostatic equilibrium at $r_{cool}$. We thus start our integration
from $r_{cool}$ towards $r_s$ and we stop when the Mach number $M=v/c_s$ is close to unity, $M\approx
0.9$. Reasonable initial values at $r_{cool}$ are: $v(r_{cool})\approx v_T \approx 10-20\, km/s$, where $v_T$ is the 
turbulent velocity and $\theta(r_{cool})$ such that $t_{cool}=\frac{5}{2}\frac{\theta}{\rho_{\scriptscriptstyle CF}\Lambda}
\le t_{Hubble}$.
In Fig.\ref{cfdynamics} we have plotted the velocity $|v(r)|$ and the isothermal sound speed $c_s(r)$
as a function of the radius $r$. This figure illustrates clearly that
the electrons are strongly accelerated in the cluster center. 
(For an extended discussion: see \citet{pm99}.) Fig.\ref{cfdensity} shows the corresponding
electron density profile $n_{CF}(r)$ in the cooling flow region.

\section{SZ contribution due to cooling flow bulk motion}

It is well known that the Kompaneets equation \citep{ko57}, resulting from a Boltzmann transport equation, describes the
spectral shift of the CMB photons due to the inverse Compton process on the 
hot electrons of the intracluster 
medium: the SZ-effect - for a review, see \citet{bi99}.
\\
The frequency dependent intensity change of the CMB photon 
field $\Delta I_K(x)$ after 
integration along the line of sight over the cluster ($cl$) dimension can be 
expressed as follows \citep{sz72,re95}: 
\begin{equation}
\Delta I_K(x)=i_0\, g(x) \int\limits_{cl}\left(\frac{kT_{e,cl}}{m_ec^2}\right) \sigma_T n_{cl}\, dl_{cl},
\label{sz1}
\end{equation}
where $x=\frac{h\nu}{kT}$ is the dimensionless frequency  with $T$ the CMB temperature and $i_0=\frac{2(kT)^3}{(hc)^2}$.
The integral is the Comptonization parameter $y_{\scriptscriptstyle K}$ describing the cluster properties with $T_{e,cl}$,\,$m_e$ the 
electron cluster
temperature and mass, respectively. $n_{cl}$ is the electron number density in the cluster and $\sigma_T$ the 
Thomson 
cross section. ($k$, $h$, $c$ are the Boltzmann constant, the Planck constant and the speed of light, respectively.)
The function $g(x)$ defines the spectral shape of the thermal SZ-effect where, for the plasma in clusters, we have
$T_{e,cl}\gg T$:
\begin{equation}
g(x)=\frac{x^4\exp x}{(\exp x-1)^2}\left( \frac{x(\exp x +1)}{\exp x-1}-4\right). \label{sz2}
\end{equation}

It is crucial to notice that the standard Kompaneets equation describes a {\it static scatterer}, 
assuming that in the average the electrons are macroscopically at rest. This is
no longer true for the electrons in an accelerated cooling flow, as it is 
described by the system of Eqs.(\ref{cf3}). 
The Kompaneets equation has 
thus to be modified in such a way, that the (macroscopic) bulk velocity 
of the 
moving electron media is explicitly taken into account.\\
\citet{ps97} gave a very detailed 
analysis of Compton scattering by static and moving media. They made a careful distinction between the
electron rest frame, the fluid frame (comoving with the fluid) and the system frame. Starting from the 
Boltzmann equation in the system frame, introducing the proper Lorentz transformations,
expanding to the appropriate orders and assuming that the velocity distribution in the fluid frame is a
relativistic Maxwellian, they end up with the zeroth moment of the radiative transfer equation with 
emission and absorption included.\\
Under the condition that the radiation field (CMB) is isotropic in the system frame, and introducing the macroscopic
electron bulk velocity $v(r)$, the {\it extended} Kompaneets equation becomes \citep{ps97}:
\begin{equation}
\frac{1}{n_e \sigma_T c}\frac{\partial I}{\partial t}=\frac{\epsilon}{m_e c^2}\frac{\partial}{\partial\epsilon}\left(
 \epsilon I\right)+\left(\frac{k T_{e,cl}}{m_e c^2}+\frac{v^2}{3 c^2}\right)\left[-4\epsilon\frac{\partial I}
{\partial\epsilon}+\epsilon\frac{\partial^2}{\partial\epsilon^2}(\epsilon I)\right], \label{sz2.1}
\end{equation}
where we neglected absorption, emission and induced scattering. $\epsilon$ is the photon energy and 
$I$ the corresponding intensity.\\
For CMB photons we have $\epsilon \ll kT_{e,cl}+\frac{1}{3}m_e v^2$. Thus, the above Eq.(\ref{sz2.1})  becomes:
\begin{equation}
\frac{1}{n_e \sigma_T c}\frac{\partial I}{\partial t}=\left(\frac{k T_{e,cl}}{m_e c^2}+\frac{v^2}{3 c^2}\right)
\left[ -2\epsilon\frac{\partial I}{\partial\epsilon}+\epsilon^2\frac{\partial^2 I}{\partial\epsilon^2}\right].
\label{sz3}
\end{equation}
Eq.(\ref{sz3}) shows that if the radiation field is isotropic in the system frame, Comptonization
of {\it the bulk motion of the electrons inside the cluster} is described entirely by second order terms  
in $v$, since all first order terms in $v$ vanish identically. 
The effect is clearly different 
from the kinematic SZ-effect, where the cluster as a whole moves through the 
CMB radiation. This equation reduces, of course, to the standard Kompaneets 
equation for $v=0$. Thus, the important point is that the effect of a 
non-zero velocity field gives an {\it additive contribution} to the standard 
thermal SZ-effect.

This way we can express the bulk motion contribution
due to the cooling flow (CF) to the SZ-effect as follows:
\begin{equation}
\frac{\partial I}{\partial t}=\left(\frac{\partial I}{\partial t}\right)_{K}+
\left(\frac{\partial I}{\partial t}\right)_{CF},                    \label{sz4}
\end{equation}
where the first term on the right hand side is given by the standard Kompaneets equation and the second term
is due to the moving electron media in the cooling flow.\\
Relating the CMB photon intensity field $I$ to the occupation number $\eta$ through $I=i_0 x^3 \eta$, we have for
the cooling flow contribution from Eq.(\ref{sz3}): 
\begin{equation}
\left(\frac{\partial I}{\partial t}\right)_{CF}=\left[-4x\frac{\partial}{\partial x}(x^3 \eta)+x\frac{\partial^2}
{\partial x^2}(x^4\eta)\right]i_0 \frac{\sigma_T n_{\scriptscriptstyle CF}}{c} \frac{v^2}{3},   \label{sz5}
\end{equation}
where $n_{\scriptscriptstyle CF}$ is the electron number density in the cooling flow region.
Integrating over the cluster cooling flow region  
and assuming a Planckian photon field for $\eta$, we find:
\begin{equation}
\Delta I_{CF}(x)=i_0\,g(x)\frac{1}{c^2}\int\limits_{\scriptscriptstyle CF} n_{\scriptscriptstyle CF}(r) \sigma_T \frac{v^2(r)}{3}\,dl_{\scriptscriptstyle CF}, 
\label{sz6} 
\end{equation}
which is the contribution due to the cooling flow bulk motion inside the cluster frame to the SZ-effect. From Eq.(\ref{sz3}) 
we see that the spectral shape $g(x)$ of this additional term
is the same as for the usual thermal SZ-effect. This clearly makes it 
more difficult to distinguish this contribution from the usual SZ-effect.

\section{Results}

In the following we get an estimate of this 
additional SZ contribution.
In order to compare bulk motion contribution in the cooling flow region
to the classical thermal SZ-effect, we introduce the following 
frequency independent ratio:
\begin{equation}
\alpha=\frac{\Delta I_{CF}} {\Delta I_K}=
\frac{y_{\scriptscriptstyle CF}} {y_{\scriptscriptstyle K}},
\label{sz7}
\end{equation}
with $y_{\scriptscriptstyle CF}=\int\limits_{\scriptscriptstyle CF}
\frac{n_{\scriptscriptstyle CF}(r)v^2(r)}{3c^2}\sigma_T\,dl_{\scriptscriptstyle CF}$ 
and $y_{\scriptscriptstyle K}=\int\limits_{cl}
\frac{k T_{e,cl}}{m_e c^2}\sigma_T n_{cl}(r)\,dl_{cl}$, the Comptonization parameter.\\
We consider a typical spherical cluster with an 
isothermal $\beta$-model for the electron
cluster density $n_{cl}$. When the line of sight goes through 
the center of a spherical cluster, we get for a large cluster radius $r_{cl}$, $r_{cl}\gg r_c$ - see e.g. \citep{pu00}:
\begin{equation}
y_{\scriptscriptstyle K}=\frac{k\, T_{e,cl}}{m_e c^2}\,\sigma_T \, r_c\, n_{0,cl}\, 
\mathcal{B}(\frac{3}{2}\beta-\frac{1}{2},
\frac{1}{2}),             \label{sz8}
\end{equation}
where $r_c$ is the core radius, $n_{0,cl}$ the central 
electron density of the cluster and 
$\mathcal{B}$ the Beta function. This way we find:
\begin{equation}
\alpha=\frac{2\, m_e}{3\,k\, T_{e,cl}\, r_c\, n_{0,cl}\, \mathcal{B}(\frac{3}{2}\beta-\frac{1}{2},\frac{1}{2})}\int\limits_{r_s}^
{r_{cool}}n_{\scriptscriptstyle CF}(r)v^2(r)\,dl_{\scriptscriptstyle CF}.    
\label{sz9}
\end{equation} 
The integral is limited to the cooling flow region from $r_{cool}$ to the sonic radius $r_s$, as explained in section 2, along 
the line of sight through the cluster center. The integration boundary $r_s$ depends on the mass deposition 
rate and the initial value conditions\footnote[2]{The sonic radius $r_s$ is typically between $1$ and $15\,kpc$.}. 
\\
As an example we take as typical values for clusters of galaxies the
numbers found for \astrobj{A426}, \astrobj{A478}, \astrobj{A1795}, \astrobj{A2029} and \astrobj{A2142} \citep{wh97}.
All these clusters have a cooling flow with a mass deposition rate of $\dot m \sim -300~ 
M_{\odot}/yr$. 
The values for $\beta$ and the central electron cluster density $n_{0,cl}$ are from \citet{mo99}. 
With these parameters we can estimate the ratio $\alpha$ between the cooling flow bulk motion 
contribution and the classical 
thermal SZ-effect, see Eq.(\ref{sz9}), which turns out to be  
of the order $10^{-5}$. 
As we cut out the most inner part of the cooling flow region, from the center to the sonic
radius $r_s$, this value must be interpreted as a lower limit to this effect.
Thus at first glance  the cooling flow bulk motion contribution is not very 
relevant for SZ observations and for the subsequent determination of the Hubble constant.
\\
Nevertheless, as $\alpha=\alpha(T_{e,cl}, n_{0,cl}, \beta, \dot{m}, r_c, r_{cool})$  depends on 
several observationally inferred parameters which can substantially change, we performed also a 
numerical investigation by letting vary the various parameters entering into the determination of 
$\alpha$. We find that, for instance:
\begin{itemize} 
\item For big mass deposition rates $\dot{m} (\approx -3000\,M_{\odot}/yr)$ and small cooling 
and core radii $(\approx 100-150\,kpc)$ $\alpha$ increases to  $10^{-4}$, by assuming that the 
other parameters have still average cluster values such as: $T_{e,cl}=7.5\,keV$, $ \beta=\frac{2}{3}$,  
$n_{0,cl}=2.2\cdot 10^{-2}cm^{-3}$. 
\vskip2mm
\item Whereas we expect $\beta\approx\frac{2}{3}$ and $T_{e,cl}\approx 7.5\,keV$ not to vary 
substantially, the value for the central electron cluster density could easily be smaller  by two 
orders of magnitude, thus reaching values as low as $n_{0,cl}\approx 10^{-4}cm^{-3}$. 
If this is 
the case then $\alpha$ can be of the order $10^{-3}-10^{-2}$. 
Notice that $n_{CF}$, since it depends on the total gravitating mass as can be seen from the system of Eqs.(\ref{cf1}),
will not change much by varying $n_{0,cl}$.
\vskip2mm
\item When taking into account the finite extension of a cluster or
an aspherical gas distribution
it turns out that $y_{\scriptscriptstyle K}$ 
gets smaller by as much as 30\%
in some cases, see \citet{pu00}. Accordingly, $\alpha$ 
increases by $\approx 40\%$.
\end{itemize}
Thus, the cooling flow bulk motion contribution to the SZ-effect can reach 
for some clusters the
percent level and become relevant for a correct
analysis of the SZ observations. Especially in view of the rapid
progress in the observational techniques - which will provide in the near
future much more accurate measurements - one might then be capable
of observing also small deviations from the standard thermal SZ-effect.
This will then enable observers either to obtain better determinations
of the Hubble constant or more details on the state of the cluster.

\section{Summary and Outlook}

We showed how the Kompaneets equation has to be modified to include the effect of 
the bulk motion due to a homogeneous steady-state cooling flow.
This effect is quadratic in the cooling flow velocity, as it is measured in the cluster frame, and thus
different from the usual kinematic SZ-effect, which is linear in the velocity. 
\\
The contribution strongly depends on the specific dynamics of the 
cooling flow and the cluster properties. Generally, it turns out to be rather
small. However, for some clusters which have more extreme parameters, though not irrealistic,
one finds a contribution at the percent level. One might speculate that
this effect could be relevant when observing young clusters which are not yet virialized, still in their formation 
phase and for which one might expect the presence of rather large and extended regions
with bulk motions, or perhaps in superclusters which are still in the formation phase.
Such young clusters might be observable in future with the Planck 
satellite.
\\
Another interesting situation might arise from shocks in merging 
processes of clusters of galaxies, in situations where the average velocity along a line of sight 
approximately vanishes. If so, the merging process will not be 
detected through the kinematic SZ-effect which is linear in the velocity,
but could be instead observed through the above discussed effect
which is quadratic in the velocity.

\noindent{\bf Acknowledgments}

We would like to thank S. Majumdar and S. Schindler for useful discussions.
This work was supported by the Swiss National Science Foundation. Part of the work of D.P. has 
been conducted under the auspices of 
the {\it D\raisebox{0.5ex}{r} Tomalla Foundation}.

\end{document}